%% file: ms.tex
\documentclass[11pt]{article}

\usepackage[letterpaper,margin=1in]{geometry}
\usepackage{amsmath}
\usepackage{amssymb}
\usepackage{booktabs}
\usepackage{graphicx}
\usepackage{url}

\input{experiment_macros.tex}

\title{WIRED: Weighted Adaptive Prediction with Structured Dependence for Probabilistic Multiseries Forecasting}
\author{Giancarlo Vercellino\\
\texttt{giancarlo.vercellino@gmail.com}}
\date{August 13, 2026}

\begin{document}
\maketitle

\begin{abstract}
This paper presents WIRED, an R package algorithm for joint probabilistic forecasting of multiple related time series. WIRED combines a library of simple marginal predictive distributions, CRPS-based adaptive mixture weights, and a Gaussian or Student $t$ copula for cross-series simulation. We evaluate the implementation in a benchmark with four synthetic data-generating processes (DGPs), three forecast horizons, 30 replicates per DGP-horizon pair, nine ablations and external baselines, and a rolling-origin study on the built-in \texttt{EuStockMarkets} data. The central contribution is architectural and diagnostic. WIRED separates adaptive marginal expert aggregation from dependence reconstruction; the benchmark supports explicit dependence modeling, but shows that the current CRPS-extrapolated softmax weighting is not yet robust enough to dominate simpler bootstrap or equal-weight alternatives. The paper therefore identifies a working layer of the design, a bottleneck in the marginal aggregation layer, and a concrete research path for more regularized probabilistic ensemble construction.
\end{abstract}


\section{Introduction}

Forecast users often need more than point forecasts. In planning, finance, supply chains, capacity management, and monitoring systems, the object of interest is a joint distribution over several future quantities. Marginal uncertainty matters, but so does dependence: simultaneous stress events, offsetting moves, and tail co-movement can change decisions even when univariate forecast accuracy is unchanged.

WIRED addresses this setting with a modular design. It treats each series and horizon as a marginal probabilistic forecasting problem, combines several simple predictive distributions using recent scoring evidence, and imposes cross-sectional coherence with an adaptive copula. The forecast-combination design follows the long-standing evidence that mixtures of simple forecasts can be more robust than selecting a single model \cite{bates1969combination,timmermann2006forecast}. The goal is pragmatic: preserve the robustness and interpretability of simple forecasters while producing simulation draws that can be used directly for scenario analysis.

This paper makes three primary contributions. First, it gives a modular architecture for multivariate probabilistic forecasting that separates marginal model uncertainty from cross-series dependence uncertainty:
\[
  \text{forecast experts}
  \rightarrow
  \text{marginal mixture}
  \rightarrow
  \text{adaptive copula}
  \rightarrow
  \text{joint scenarios}.
\]
Second, it evaluates dependence reconstruction as a separate forecasting component by comparing structured copula draws with independent sampling of the same WIRED marginals. Third, it reports a deliberately mixed ablation study showing that the dependence layer is useful, while the current adaptive marginal weighting rule requires stronger regularization or bootstrap-aware alternatives.

\section{Problem Setup}

Let $\{y_t\}_{t=1}^T$ be a $p$-dimensional positive level process, where $y_t=(y_{t1},\ldots,y_{tp})^\top$. For a horizon $h$, WIRED models a transformed future movement
\begin{equation}
  x_{t+h,j} =
  \begin{cases}
    y_{t+h,j} - y_{t,j}, & \text{additive},\\
    y_{t+h,j}/y_{t,j} - 1, & \text{multiplicative},\\
    \log(y_{t+h,j}/y_{t,j}), & \text{log-multiplicative}.
  \end{cases}
\end{equation}
The fitted object provides samplers and distribution functions on the transformed scale and, when requested, on the level scale. For a fixed horizon, the target is a joint predictive distribution
\begin{equation}
  F_h(x_1,\ldots,x_p \mid y_{1:T})
\end{equation}
with useful marginal calibration and a dependence structure suitable for simulation.

\section{Algorithm}

\noindent\fbox{%
\begin{minipage}{0.96\linewidth}
\textbf{Canonical WIRED procedure}

\smallskip
\noindent\textbf{Inputs.} A $p$-variate level panel $y_{1:T}$; forecast horizon(s) $h\in H$; number of joint draws $M$; transformation mode; marginal expert library; CRPS backtest settings; dependence metric; copula and correlation-adaptation controls.

\smallskip
\noindent\textbf{Outputs.} Marginal predictive distributions; joint transformed-scale and level-scale draws; adaptive mixture weights; dependence metadata; diagnostic tables and plots.

\begin{enumerate}
  \item \textbf{Transform.} For each horizon $h$, convert levels into horizon-aligned movements $x_{t+h,j}$ on the additive, multiplicative, or log-multiplicative scale.
  \item \textbf{Fit marginals.} For each series and horizon, fit the candidate forecasting library and estimate recent CRPS histories by expanding-window backtests.
  \item \textbf{Weight experts.} Extrapolate each candidate's next-window CRPS and convert the predicted scores into temperature-scaled mixture weights.
  \item \textbf{Estimate dependence.} Align transformed historical moves, estimate a static, rolling, EWMA, or regime-adaptive correlation matrix, then shrink and repair it for copula simulation.
  \item \textbf{Sample scenarios.} Draw uniforms from the Gaussian or Student $t$ copula, map them through marginal mixture quantiles, and invert the transformation to return coherent level paths.
\end{enumerate}

\[
  y_{1:T}\rightarrow \{x_{t+h,j}\}\rightarrow \{F_{j,h}\}
  \rightarrow C_h \rightarrow \{X_h^{(m)}\}_{m=1}^M
  \rightarrow \{\tilde{y}_{T+h}^{(m)}\}_{m=1}^M .
\]
\end{minipage}}

\begin{table}[ht]
\centering
\caption{Core WIRED notation.}
\label{tab:wired-notation}
\begin{tabular}{p{0.16\linewidth}p{0.76\linewidth}}
\toprule
Symbol & Meaning\\
\midrule
$y_{1:T}$ & Observed multivariate level history through time $T$.\\
$y_t$ & $p$-dimensional level vector observed at time $t$.\\
$p$ & Number of component series in the panel.\\
$h\in H$ & Forecast horizon and set of horizons requested by the user.\\
$M$ & Number of joint predictive draws to simulate.\\
$x_{t+h,j}$ & Horizon-$h$ transformed movement for series $j$.\\
$F_{j,h}$ & Marginal predictive distribution for series $j$ at horizon $h$.\\
$C_h$ & Gaussian or Student $t$ copula used to couple the horizon-$h$ marginals.\\
$X_h^{(m)}$ & Simulated transformed-scale joint draw $m$ at horizon $h$.\\
$\tilde{y}_{T+h}^{(m)}$ & Simulated level-scale scenario $m$ for time $T+h$.\\
\bottomrule
\end{tabular}
\end{table}

\section{Algorithmic Components}

\subsection{Marginal Forecast Library}

For each series $j$ and horizon $h$, WIRED constructs a candidate library of eight predictive distributions:
\begin{enumerate}
  \item a naive PERT distribution centered on the most recent transformed move;
  \item an automatic ARIMA forecast distribution \cite{hyndman2008automatic};
  \item an exponentially weighted moving-average Gaussian forecaster;
  \item a historical bootstrap over horizon-aligned moves;
  \item a drift plus residual bootstrap;
  \item a volatility-scaled naive Gaussian forecaster;
  \item a robust median/MAD forecaster with Laplace or Gaussian shape; and
  \item a shrunk quantile-regression forecaster \cite{koenker1978regression}.
\end{enumerate}
Each candidate exposes a random sampler $r(\cdot)$, density $f(\cdot)$, cumulative distribution $F(\cdot)$, and quantile function $Q(\cdot)$.

\subsection{CRPS-Weighted Mixture}

The continuous ranked probability score is a strictly proper scoring rule for univariate predictive distributions \cite{matheson1976scoring,gneiting2007strictly}. For a sample forecast $X_1,\ldots,X_m$ and realization $z$, WIRED estimates CRPS as
\begin{equation}
  \widehat{\mathrm{CRPS}}(X,z)
  =
  \frac{1}{m}\sum_{i=1}^m |X_i-z|
  -
  \frac{1}{2m^2}\sum_{i=1}^m\sum_{k=1}^m |X_i-X_k|.
\end{equation}
For each candidate forecaster, expanding-window backtests produce a recent CRPS series. A robust Theil-Sen style slope estimate extrapolates the next-window score, producing predicted scores $\hat{s}_{1:K}$. WIRED converts these predicted scores into mixture weights
\begin{equation}
  w_k = \frac{\exp(-\hat{s}_k/\tau)}{\sum_{\ell=1}^K \exp(-\hat{s}_\ell/\tau)},
\end{equation}
where $\tau$ is a data-adaptive temperature, defaulting to the empirical standard deviation of the predicted scores. This rule has the same exponential-weight form used in finite-expert prediction \cite{cesabianchi2006prediction}, but here the losses are predicted proper scores rather than accumulated realized losses.

\noindent The same weights can be read as the solution of a simple entropy-regularized linear loss problem. If $\Delta_K=\{w\in\mathbb{R}^K_+:\sum_{k=1}^K w_k=1\}$ and the predicted losses $\hat{s}_{1:K}$ are fixed, then
\begin{equation}
  w =
  \arg\min_{v\in\Delta_K}
  \left\{
    \sum_{k=1}^K v_k\hat{s}_k
    +
    \tau\sum_{k=1}^K v_k\log v_k
  \right\}.
\end{equation}
The first term favors components with lower predicted CRPS, while the entropy term keeps weights from collapsing too sharply. This interpretation is useful but modest: the empirical issue is whether the predicted scores are accurate enough for the regularized softmax rule to improve on simpler alternatives.

The marginal predictive distribution is
\begin{equation}
  F_j(x) = \sum_{k=1}^K w_k F_{jk}(x).
\end{equation}
Mixture quantiles are computed by monotone numerical inversion of the mixture CDF.

\subsection{Structured Dependence}

After fitting marginal mixtures, WIRED estimates a correlation prototype from aligned transformed histories. Copulas provide a convenient way to separate marginal predictive calibration from cross-series dependence modeling \cite{nelsen2006introduction,genest2007copula}. Let $Z_h$ be the matrix of aligned historical horizon-$h$ movements used for dependence estimation. The user may choose Pearson correlation or rank correlations mapped to the latent elliptical correlation scale:
\begin{equation}
  \widehat R_{ij} =
  \begin{cases}
    \widehat\rho^{P}_{ij}, & \text{Pearson},\\
    \sin(\pi\widehat\tau_{ij}/2), & \text{Kendall},\\
    2\sin(\pi\widehat\rho^{S}_{ij}/6), & \text{Spearman},
  \end{cases}
\end{equation}
where $\widehat\rho^{P}_{ij}$, $\widehat\tau_{ij}$, and $\widehat\rho^{S}_{ij}$ are the empirical Pearson, Kendall, and Spearman correlations of columns $i$ and $j$ of $Z_h$. This mapping is the standard conversion from rank dependence to the Gaussian or Student $t$ copula correlation parameter. The Student $t$ option is motivated by dependence modeling in settings where joint tail behavior matters \cite{patton2006modelling}. The implementation supports four adaptation modes:
\begin{itemize}
  \item static: estimate one correlation matrix from all aligned history;
  \item rolling: estimate correlation over the latest window;
  \item EWMA: update second moments with exponential decay; and
  \item regime: blend calm and stress correlation matrices using a smooth stress score.
\end{itemize}
In regime mode, WIRED forms a smoothed stress score $s_t$ from recent average absolute transformed moves, estimates $R_{\mathrm{calm}}$ from low-stress rows and $R_{\mathrm{stress}}$ from high-stress rows, and blends them as
\begin{equation}
  R_{\mathrm{reg}}
  =
  (1-\omega_T)R_{\mathrm{calm}}+\omega_T R_{\mathrm{stress}},
  \qquad
  \omega_T =
  \{1+\exp[-k(s_T-c)]\}^{-1},
\end{equation}
where $c$ is the midpoint between the calm and stress stress-score thresholds and $k$ controls the sharpness of the transition. The selected matrix is then shrunk toward the identity,
\begin{equation}
  R_\alpha=(1-\alpha)\widehat R+\alpha I_p,
\end{equation}
and repaired by flooring very small eigenvalues before rescaling to a unit-diagonal correlation matrix. A Gaussian or Student $t$ copula generates uniforms $U_{1:p}$, which are mapped through marginal quantile functions:
\begin{equation}
  X_j = Q_j(U_j), \qquad j=1,\ldots,p.
\end{equation}
The same uniforms can be mapped through transformed-scale and level-scale quantile functions, giving coherent scenario draws.

\section{Software Interface}

The package exports a single main function:
\begin{verbatim}
wired(ts_set, future, dates = NULL,
      mode = c("additive", "multiplicative", "log_multiplicative"),
      n_testing = 30, dep_metric = c("kendall", "spearman", "pearson"),
      corr_adapt = c("static", "ewma", "rolling", "regime"),
      copula = c("gaussian", "t"), ...)
\end{verbatim}
The return value includes per-horizon fitted components, joint samplers on transformed and level scales, a recorded base R forecast plot, and metadata describing dependence and copula controls. A minimal workflow is:
\begin{verbatim}
set.seed(1)
ts_set <- data.frame(
  A = 100 + cumsum(rnorm(240, 0, 1)),
  B =  90 + cumsum(rnorm(240, 0, 1)),
  C =  80 + cumsum(rnorm(240, 0, 1))
)

fit <- wired(ts_set, future = 3, mode = "additive",
             corr_adapt = "rolling", copula = "t",
             n_testing = 2, n_crps_mc = 50, q_grid_size = 100)

draws <- fit$rfun_level(512)  # array: horizon x draw x series
str(fit$meta)
\end{verbatim}
The package imports \texttt{forecast}, \texttt{quantreg}, \texttt{MASS}, \texttt{mc2d}, and \texttt{imputeTS}; it is licensed under GPL-3.

\section{Benchmark Design}

\subsection{Synthetic Data-Generating Processes}

The synthetic benchmark uses four positive-level multiplicative-growth data-generating processes (DGPs): \texttt{regime\_heavy\_tail}, \texttt{static\_gaussian}, \texttt{break\_correlation}, and \texttt{independent\_series}. Each replicate contains $p=4$ series and \TrainLength{} training observations. Let
\[
  \mu=(0.00035,0.00020,0.00010,-0.00005)^\top,\qquad
  y_1=(100,90,80,110)^\top .
\]
All four DGPs evolve as
\begin{equation}
  y_{t,j}=y_{t-1,j}(1+g_{t,j}),
\end{equation}
where $g_{t,j}$ is clipped below at $-0.20$ only to avoid impossible negative levels. The growth laws are
\begin{align*}
\texttt{static\_gaussian}:&\quad
  g_t \sim N(\mu, D_s R_s D_s),
  \quad D_s=\operatorname{diag}(0.0060,0.0050,0.0045,0.0040),\\
&\quad (R_s)_{ij}=0.45 \text{ for } i\ne j;\\
\texttt{independent\_series}:&\quad
  g_{t,j}=\mu_j+\sigma_j\epsilon_{t,j},
  \quad \epsilon_{t,j}\overset{\mathrm{iid}}{\sim} t_7,
  \quad \sigma=(0.0060,0.0050,0.0045,0.0040);\\
\texttt{break\_correlation}:&\quad
  g_t=\mu+0.0006\sin(2\pi t/36)\mathbf{1}+z_t,
  \quad z_t\sim N(0,D_qR_qD_q),
\end{align*}
where $q=\mathrm{pre}$ until $t>\lfloor0.72T\rfloor$ and $q=\mathrm{post}$ afterwards. In this DGP,
\[
D_{\mathrm{pre}}=\operatorname{diag}(0.0045,0.0040,0.0038,0.0035),\quad
D_{\mathrm{post}}=\operatorname{diag}(0.0090,0.0080,0.0070,0.0065),
\]
\[
R_{\mathrm{pre}}=
\begin{pmatrix}
1&0.10&-0.05&0.05\\
0.10&1&0.12&-0.10\\
-0.05&0.12&1&0.08\\
0.05&-0.10&0.08&1
\end{pmatrix},
\qquad
(R_{\mathrm{post}})_{ij}=0.70 \text{ for } i\ne j .
\]
Finally, \texttt{regime\_heavy\_tail} uses a Markov stress state $S_t\in\{0,1\}$ with $\Pr(S_t=1\mid S_{t-1}=0)=0.08$ and $\Pr(S_t=1\mid S_{t-1}=1)=0.65$:
\[
  g_t=\mu+0.0008\sin(2\pi t/48)\mathbf{1}
  +a_{S_t}\lambda_{S_t}F_t+b_{S_t}\epsilon_t,
\]
where $F_t\sim t_{\nu_{S_t}}$, $\epsilon_{t,j}\overset{\mathrm{iid}}{\sim}t_{\nu_{S_t}+2}$, $\nu_0=9$, $\nu_1=4$, $(a_0,b_0)=(0.0035,0.0015)$, $(a_1,b_1)=(0.011,0.0040)$, $\lambda_0=(0.65,0.45,0.15,-0.25)^\top$, and $\lambda_1=(1.15,0.95,0.70,0.40)^\top$. Table~\ref{tab:config} gives the benchmark configuration.

\begin{table}[t]
\centering
\caption{Benchmark configuration.}
\label{tab:config}
\resizebox{\linewidth}{!}{\input{table_experiment_config.tex}}
\end{table}

\subsection{Forecast Variants}

The benchmark compares nine variants:
\begin{itemize}
  \item \textbf{WIRED-full}: CRPS-weighted WIRED marginals, regime-adaptive rank dependence, shrinkage repair, and Student $t$ copula;
  \item \textbf{No tail copula}: same marginals and correlation matrix, but Gaussian copula;
  \item \textbf{Static dependence}: same marginals and Student $t$ copula, but static correlation;
  \item \textbf{Rolling dependence}: same marginals and Student $t$ copula, but rolling correlation;
  \item \textbf{Independent marginals}: same WIRED marginals, sampled independently;
  \item \textbf{Equal weights}: same candidate library and full dependence layer, but uniform marginal mixture weights;
  \item \textbf{ARIMA independent}: independent automatic ARIMA marginals;
  \item \textbf{Bootstrap independent}: independent historical-bootstrap marginals; and
  \item \textbf{Gaussian copula bootstrap}: historical-bootstrap marginals coupled by a static Gaussian copula.
\end{itemize}

\subsection{Real-Data Rolling-Origin Study}

As a small real-data check, we use the built-in R \texttt{EuStockMarkets} data, containing daily DAX, SMI, CAC, and FTSE levels. Forecasts are evaluated at eight rolling origins and horizons 1, 5, and 20. This benchmark is intentionally modest, but it provides a non-synthetic check without requiring external data downloads.

\subsection{Scoring Rules}

Let $X_1,\ldots,X_m$ be joint forecast draws and $y$ the realized vector. The sample energy score is a multivariate generalization of CRPS for ensemble-style predictive samples \cite{gneiting2008assessing}:
\begin{equation}
  \widehat{\mathrm{ES}}(X,y)
  =
  \frac{1}{m}\sum_{i=1}^m \|X_i-y\|_2
  -
  \frac{1}{2m^2}\sum_{i=1}^m\sum_{k=1}^m \|X_i-X_k\|_2.
\end{equation}
The variogram score of order $r=1/2$ is estimated as a dependence-sensitive proper score based on pairwise component differences \cite{scheuerer2015variogram}:
\begin{equation}
  \widehat{\mathrm{VS}}(X,y)
  =
  \sum_{a<b}
  \left(
    |y_a-y_b|^r
    -
    \frac{1}{m}\sum_{i=1}^m |X_{ia}-X_{ib}|^r
  \right)^2.
\end{equation}
We also report average marginal CRPS, empirical 80\% interval coverage, and average 80\% interval width. Lower CRPS, energy score, and variogram score are better. Coverage and width are interpreted jointly.

\section{Results}

\subsection{Synthetic Benchmark}

\begin{table}[t]
\centering
\caption{Synthetic benchmark results across \NumDGPs{} DGPs, \NumHorizons{} horizons, and \NumReplicates{} replicates per DGP-horizon pair. Values are means $\pm$ standard deviations.}
\label{tab:synthetic}
\resizebox{\linewidth}{!}{\input{table_metrics.tex}}
\end{table}

\begin{table}[t]
\centering
\caption{Paired score differences versus WIRED-full on the synthetic benchmark. Negative values improve on WIRED-full. Intervals are approximate 95\% standard-error intervals across paired DGP-horizon-replicate cases.}
\label{tab:deltas}
\resizebox{\linewidth}{!}{\input{table_deltas.tex}}
\end{table}

Table~\ref{tab:synthetic} shows that WIRED-full is not the overall winner in this benchmark. The Gaussian-copula bootstrap baseline has the best average CRPS, energy score, and variogram score. Table~\ref{tab:deltas} shows two different effects. First, replacing WIRED's copula layer with independent sampling worsens the variogram score by \IndependentVariogramDelta{} on average, confirming that dependence modeling matters for the WIRED marginals. Second, the equal-weight and bootstrap baselines often improve marginal and joint scores, indicating that the current adaptive CRPS weighting is not uniformly beneficial under these benchmark settings.

\begin{figure}[t]
\centering
\includegraphics[width=0.95\linewidth]{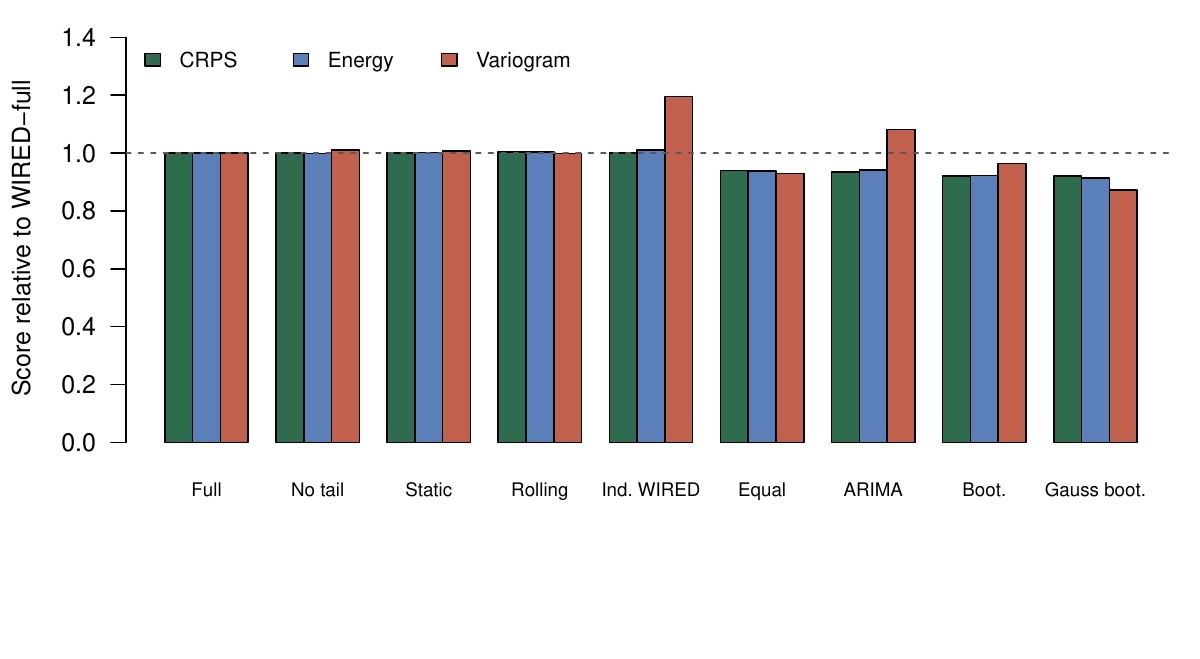}
\caption{Synthetic benchmark scores normalized to WIRED-full. Values above one are worse than WIRED-full; values below one are better.}
\label{fig:score-ablation}
\end{figure}

\begin{figure}[t]
\centering
\includegraphics[width=0.95\linewidth]{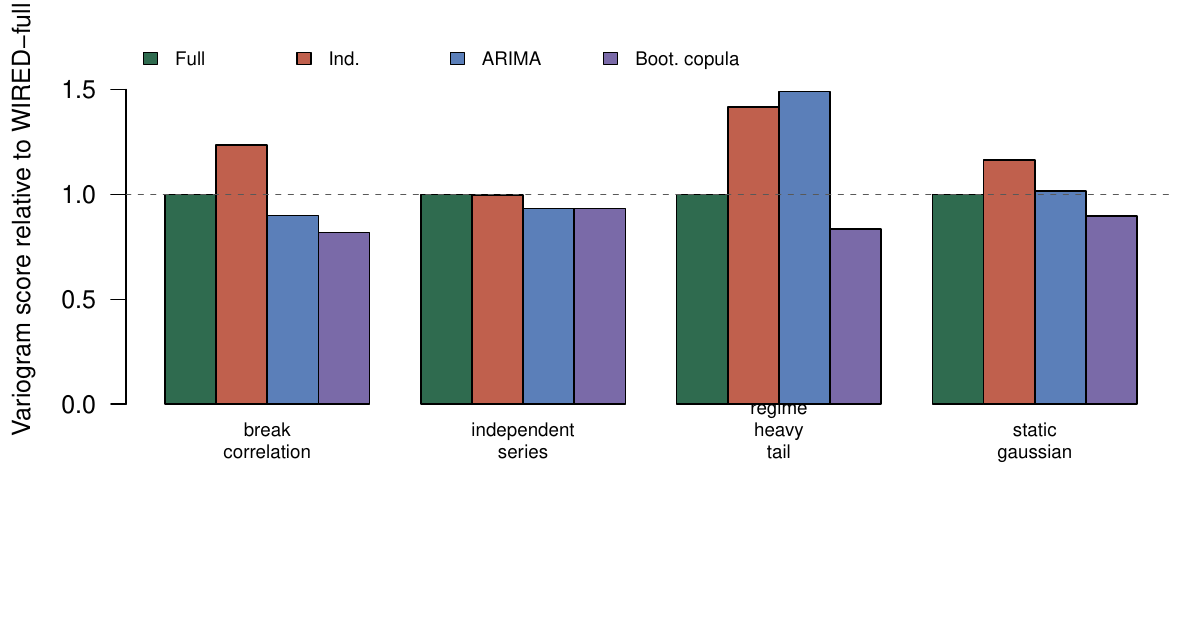}
\caption{Synthetic variogram scores by DGP, normalized to WIRED-full.}
\label{fig:score-by-dgp}
\end{figure}

\begin{table}[t]
\centering
\caption{Selected synthetic results by DGP.}
\label{tab:by-dgp}
\resizebox{\linewidth}{!}{\input{table_by_dgp.tex}}
\end{table}

Figure~\ref{fig:score-by-dgp} and Table~\ref{tab:by-dgp} show that the independent WIRED-marginal variant is consistently worse than WIRED-full on dependence-sensitive variogram scores in correlated settings. However, Gaussian-copula bootstrap remains highly competitive and often better, especially when historical transformed moves are already a strong approximation to the next-horizon distribution.

\begin{figure}[t]
\centering
\includegraphics[width=0.95\linewidth]{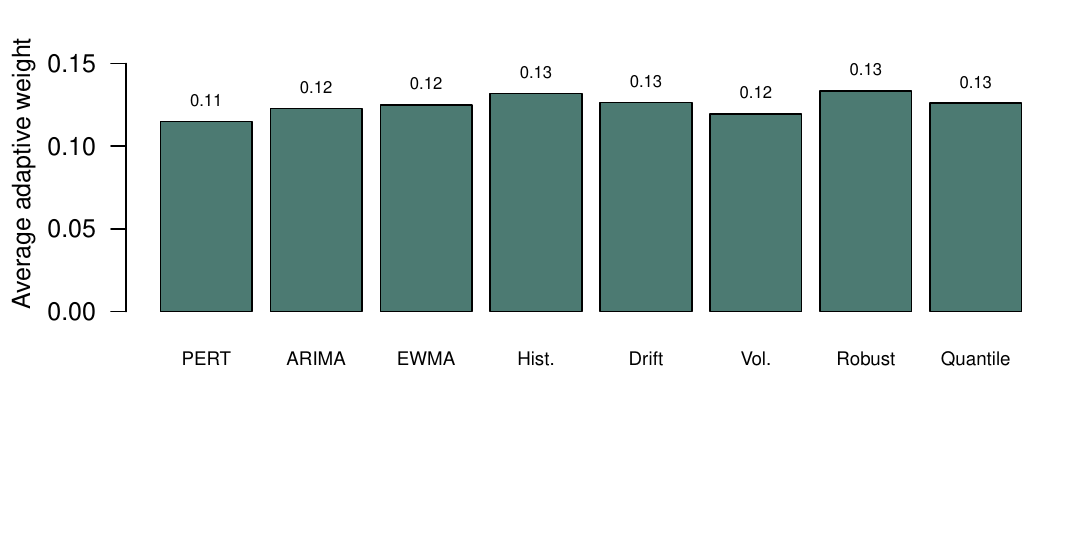}
\caption{Average adaptive marginal mixture weights across synthetic replicates, DGPs, horizons, and series.}
\label{fig:weights}
\end{figure}

Figure~\ref{fig:weights} shows that the adaptive mixture does not collapse to a single forecaster. This diversification is useful operationally, but the benchmark suggests that the current score-to-weight transformation should be calibrated more carefully.

\subsection{Real-Data Rolling-Origin Results}

\begin{table}[t]
\centering
\caption{Rolling-origin results on \texttt{EuStockMarkets}. Values are means $\pm$ standard deviations across eight origins and horizons 1, 5, and 20.}
\label{tab:real}
\resizebox{\linewidth}{!}{\input{table_real_data.tex}}
\end{table}

The real-data results in Table~\ref{tab:real} reinforce the synthetic conclusion. WIRED-full improves substantially over independently sampled WIRED marginals on the variogram score, but the Gaussian-copula bootstrap baseline again obtains the best average scores. This does not invalidate WIRED's package design; rather, it identifies a strong baseline that future versions should incorporate more directly.

\clearpage

\section{Discussion}

The benchmark supports a layerwise interpretation of WIRED. The dependence layer is useful: independent sampling of the same WIRED marginals generally harms dependence-sensitive scoring. This supports the architectural premise that good marginal probabilistic forecasts are not sufficient for good multivariate scenarios. The adaptive marginal mixture, however, is not yet clearly superior to simpler historical-bootstrap and equal-weight alternatives. Two mechanisms may explain this. First, short expanding-window CRPS histories can be noisy, especially for long horizons. Second, the softmax rule may amplify small predicted score differences when those predictions are uncertain, while historical bootstrap marginals are naturally robust in the simulated and equity-index settings considered here.

These results suggest concrete improvements. WIRED should expose stronger controls for the score-to-weight temperature, consider regularizing weights toward equal weights or a bootstrap prior, and include the Gaussian-copula bootstrap as an explicit baseline or fallback mode. More broadly, future work should treat expert-skill forecasting as its own modeling problem: the dependence reconstruction layer appears empirically useful, while the marginal aggregation layer needs better predictions of when each expert is about to be useful. The package would also benefit from benchmark helpers so users can compare WIRED against simpler alternatives on their own data before relying on the adaptive mixture.

\section{Limitations}

The synthetic benchmark is limited to four DGP families and positive multiplicative levels. The real-data study uses one built-in equity-index dataset and only eight origins. The paper uses capped simulation settings for runtime, including reduced ARIMA simulation draws and bootstrap sample sizes in the benchmark driver. Finally, copula dependence is estimated from transformed historical moves and does not model a full latent time-varying dependence process.

\section{Reproducibility}

At the time of submission, the released WIRED implementation is available from CRAN \cite{wiredcran2026}:
\begin{center}
\url{https://CRAN.R-project.org/package=wired}
\end{center}
It can be installed with \texttt{install.packages("wired")}. The experiment driver and precomputed CSV outputs are distributed as expanded arXiv ancillary files under \path{anc/paper/}, with instructions in \path{anc/README_REPRODUCE.txt}. The driver is \path{anc/paper/experiment/run_experiment.R}; it uses local package source when available and otherwise loads the installed CRAN package. The script has a smoke-test mode and a paper-benchmark mode:
\begin{verbatim}
WIRED_BENCH_MODE=quick Rscript anc/paper/experiment/run_experiment.R
WIRED_BENCH_MODE=paper Rscript anc/paper/experiment/run_experiment.R
\end{verbatim}
It writes CSV outputs, tables, and figures into the corresponding paper directories. The reported computations were run with R 4.5.1 on Windows 11.

\section{Conclusion}

WIRED provides a practical R implementation for multiseries probabilistic forecasting when users need calibrated marginal uncertainty and coherent joint scenario draws. Its strongest academic contribution is a modular design that separates adaptive marginal expert aggregation from copula-based dependence reconstruction. The benchmark supports the dependence component: structured joint sampling improves dependence-sensitive behavior relative to independent WIRED sampling. It also gives a useful negative result: the current CRPS-extrapolated marginal weighting layer does not consistently beat simpler bootstrap and equal-weight alternatives. The most useful conclusion is therefore constructive: WIRED is a promising extensible package design, and its next methodological gains should come from better expert-skill forecasting, stronger weight regularization, explicit bootstrap-copula baselines, and broader empirical validation.

\end{document}

%% file: experiment_macros.tex
\newcommand{\NumReplicates}{30}
\newcommand{\NumDGPs}{4}
\newcommand{\NumHorizons}{3}

\newcommand{\TrainLength}{240}

\newcommand{\IndependentVariogramDelta}{0.00186}

%% file: table_experiment_config.tex
\begin{tabular}{ll}
\toprule
Setting & Value \\
\midrule
Mode & paper \\
Synthetic DGPs & regime\_heavy\_tail; static\_gaussian; break\_correlation; independent\_series \\
Replicates per DGP-horizon & 30 \\
Horizons & 1, 3, 6 \\
Real-data origins/horizons & 8 origins; 1, 5, 20 \\
Training length & 240 \\
Series dimension & 4 \\
Evaluation draws & 384 \\
Forecast draws & 512 \\
Expanding-window CRPS points & 2 \\
CRPS Monte Carlo draws & 64 \\
Mixture quantile grid & 140 \\
\bottomrule
\end{tabular}

%% file: table_metrics.tex
\begin{tabular}{lccccc}
\toprule
Variant & CRPS $\downarrow$ & Energy $\downarrow$ & Variogram $\downarrow$ & 80\% cov. & 80\% width \\
\midrule
WIRED-full & 0.00650 $\pm$ 0.00334 & 0.01497 $\pm$ 0.00755 & 0.00955 $\pm$ 0.00631 & 0.757 $\pm$ 0.086 & 0.0251 $\pm$ 0.0107 \\
No tail copula & 0.00650 $\pm$ 0.00332 & 0.01496 $\pm$ 0.00749 & 0.00964 $\pm$ 0.00629 & 0.756 $\pm$ 0.086 & 0.0251 $\pm$ 0.0107 \\
Static dependence & 0.00651 $\pm$ 0.00334 & 0.01499 $\pm$ 0.00758 & 0.00961 $\pm$ 0.00637 & 0.757 $\pm$ 0.086 & 0.0251 $\pm$ 0.0107 \\
Rolling dependence & 0.00652 $\pm$ 0.00340 & 0.01502 $\pm$ 0.00770 & 0.00953 $\pm$ 0.00642 & 0.756 $\pm$ 0.087 & 0.0251 $\pm$ 0.0107 \\
Independent marginals & 0.00650 $\pm$ 0.00332 & 0.01515 $\pm$ 0.00766 & 0.01141 $\pm$ 0.00806 & 0.756 $\pm$ 0.086 & 0.0251 $\pm$ 0.0107 \\
Equal weights & 0.00611 $\pm$ 0.00269 & 0.01404 $\pm$ 0.00602 & 0.00887 $\pm$ 0.00370 & 0.789 $\pm$ 0.082 & 0.0262 $\pm$ 0.0103 \\
ARIMA independent & 0.00607 $\pm$ 0.00264 & 0.01409 $\pm$ 0.00598 & 0.01033 $\pm$ 0.00512 & 0.761 $\pm$ 0.113 & 0.0242 $\pm$ 0.0092 \\
Bootstrap independent & 0.00599 $\pm$ 0.00266 & 0.01380 $\pm$ 0.00592 & 0.00920 $\pm$ 0.00375 & 0.733 $\pm$ 0.113 & 0.0220 $\pm$ 0.0089 \\
Gaussian copula bootstrap & \textbf{0.00599 $\pm$ 0.00266} & \textbf{0.01367 $\pm$ 0.00587} & \textbf{0.00833 $\pm$ 0.00336} & 0.734 $\pm$ 0.113 & 0.0220 $\pm$ 0.0090 \\
\bottomrule
\end{tabular}

%% file: table_deltas.tex
\begin{tabular}{lccc}
\toprule
Variant & $\Delta$ CRPS & $\Delta$ Energy & $\Delta$ Variogram \\
\midrule
No tail copula & -0.00000 $\pm$ 0.00001 & -0.00001 $\pm$ 0.00003 & +0.00009 $\pm$ 0.00004 \\
Static dependence & +0.00000 $\pm$ 0.00001 & +0.00002 $\pm$ 0.00003 & +0.00006 $\pm$ 0.00004 \\
Rolling dependence & +0.00002 $\pm$ 0.00002 & +0.00004 $\pm$ 0.00004 & -0.00001 $\pm$ 0.00005 \\
Independent marginals & +0.00000 $\pm$ 0.00001 & +0.00017 $\pm$ 0.00004 & +0.00186 $\pm$ 0.00030 \\
Equal weights & -0.00039 $\pm$ 0.00015 & -0.00093 $\pm$ 0.00034 & -0.00068 $\pm$ 0.00041 \\
ARIMA independent & -0.00043 $\pm$ 0.00017 & -0.00088 $\pm$ 0.00039 & +0.00078 $\pm$ 0.00064 \\
Bootstrap independent & -0.00051 $\pm$ 0.00017 & -0.00117 $\pm$ 0.00039 & -0.00035 $\pm$ 0.00055 \\
Gaussian copula bootstrap & -0.00051 $\pm$ 0.00017 & -0.00131 $\pm$ 0.00039 & -0.00122 $\pm$ 0.00053 \\
\bottomrule
\end{tabular}

%% file: table_by_dgp.tex
\begin{tabular}{llccc}
\toprule
DGP & Variant & CRPS & Energy & Variogram \\
\midrule
break\_correlation & WIRED-full & 0.00857 & 0.01924 & 0.00950 \\
 & Independent marginals & 0.00857 & 0.01952 & 0.01173 \\
 & ARIMA independent & 0.00797 & 0.01807 & 0.00854 \\
 & Gaussian copula bootstrap & 0.00794 & 0.01762 & 0.00777 \\
\addlinespace
independent\_series & WIRED-full & 0.00608 & 0.01454 & 0.01153 \\
 & Independent marginals & 0.00607 & 0.01449 & 0.01150 \\
 & ARIMA independent & 0.00572 & 0.01362 & 0.01076 \\
 & Gaussian copula bootstrap & 0.00567 & 0.01345 & 0.01076 \\
\addlinespace
regime\_heavy\_tail & WIRED-full & 0.00620 & 0.01407 & 0.00965 \\
 & Independent marginals & 0.00620 & 0.01442 & 0.01366 \\
 & ARIMA independent & 0.00577 & 0.01335 & 0.01438 \\
 & Gaussian copula bootstrap & 0.00552 & 0.01246 & 0.00805 \\
\addlinespace
static\_gaussian & WIRED-full & 0.00515 & 0.01204 & 0.00752 \\
 & Independent marginals & 0.00517 & 0.01216 & 0.00876 \\
 & ARIMA independent & 0.00483 & 0.01132 & 0.00764 \\
 & Gaussian copula bootstrap & 0.00483 & 0.01114 & 0.00674 \\
\addlinespace
\bottomrule
\end{tabular}

%% file: table_real_data.tex
\begin{tabular}{lccccc}
\toprule
Variant & CRPS $\downarrow$ & Energy $\downarrow$ & Variogram $\downarrow$ & 80\% cov. & 80\% width \\
\midrule
WIRED-full & 0.01647 $\pm$ 0.02070 & 0.03571 $\pm$ 0.04146 & 0.01384 $\pm$ 0.01175 & 0.792 $\pm$ 0.335 & 0.0610 $\pm$ 0.0340 \\
No tail copula & 0.01644 $\pm$ 0.02052 & 0.03564 $\pm$ 0.04093 & 0.01466 $\pm$ 0.01255 & 0.792 $\pm$ 0.335 & 0.0615 $\pm$ 0.0349 \\
Static dependence & 0.01650 $\pm$ 0.02061 & 0.03572 $\pm$ 0.04127 & 0.01322 $\pm$ 0.00996 & 0.802 $\pm$ 0.338 & 0.0617 $\pm$ 0.0357 \\
Rolling dependence & 0.01663 $\pm$ 0.02082 & 0.03589 $\pm$ 0.04166 & 0.01365 $\pm$ 0.01045 & 0.771 $\pm$ 0.329 & 0.0610 $\pm$ 0.0343 \\
Independent marginals & 0.01666 $\pm$ 0.02093 & 0.03701 $\pm$ 0.04272 & 0.02472 $\pm$ 0.01911 & 0.771 $\pm$ 0.337 & 0.0615 $\pm$ 0.0351 \\
Equal weights & 0.01596 $\pm$ 0.02093 & 0.03459 $\pm$ 0.04188 & 0.01402 $\pm$ 0.01248 & 0.823 $\pm$ 0.342 & 0.0647 $\pm$ 0.0359 \\
ARIMA independent & 0.01471 $\pm$ 0.01834 & 0.03294 $\pm$ 0.03763 & 0.02275 $\pm$ 0.01853 & 0.812 $\pm$ 0.315 & 0.0605 $\pm$ 0.0340 \\
Bootstrap independent & 0.01429 $\pm$ 0.01752 & 0.03195 $\pm$ 0.03597 & 0.02168 $\pm$ 0.01960 & 0.802 $\pm$ 0.313 & 0.0578 $\pm$ 0.0342 \\
Gaussian copula bootstrap & \textbf{0.01427 $\pm$ 0.01741} & \textbf{0.03101 $\pm$ 0.03485} & \textbf{0.01288 $\pm$ 0.01012} & 0.802 $\pm$ 0.313 & 0.0576 $\pm$ 0.0343 \\
\bottomrule
\end{tabular}